\documentclass{eas}

\begin{document}
\title{Comparison of the atmosphere above the South Pole, Dome C and
  Dome A: first attempt}
\runningtitle{Hagelin \etal: Comparison of the atmosphere above Antarctica}
\author{Susanna Hagelin}
\address{INAF - Osservatorio Astrofisico di Arcetri, Largo Enrico
  Fermi 5, I-50125 Firenze, Italy \\ \email{hagelin@arcetri.astro.it; masciadri@arcetri.astro.it} }
\secondaddress{Uppsala University, Department of Earth Sciences,
  Villav\"agen 16, S-752 36 Uppsala, Sweden}
\author{Elena Masciadri}
\sameaddress{1}
\author{Franck Lascaux}
\sameaddress{1}
\author{Jeff Stoesz}
\sameaddress{1}

\begin{abstract}
The atmospheric properties above three sites (Dome A, Dome C and the
South Pole) are investigated for astronomical applications using the
monthly median of the analyses from the ECMWF (European Centre for
Medium-Range Weather Forecasts). Radiosoundings extended on a yearly
time-scale at the South Pole and Dome C are used to quantify the
reliability of the ECMWF analyses in the free atmosphere as well as
in the boundary and surface layers, and to characterize the median
wind speed in the first 100 m above the two sites. Thermodynamic
instabilities in the free atmosphere above the three sites are
quantified with monthly median values of the Richardson number. We
will present a ranking of the sites with respect to the thermodynamic
stability, using the Richardson number, and with respect to the wind
speed, in the free atmosphere (using ECMWF analyses) as well as in the
surface layer (using radiosoundings).
\end{abstract}

\maketitle

\section{Introduction}

The summits of the Internal Antarctic Plateau might be among the best
sites in the world for astronomical facilities. The free atmosphere
has low amounts of water vapour and the turbulence is concentrated to
a thin surface layer. The largest source of the turbulence in the
surface layer is the near surface winds, that are triggered by the
sloping terrain in combination with the temperature inversion.
For a more extensive analysis we refer the reader to  Hagelin {\em et al.\/}
(\cite{hag08a}), in this contribution we briefly summarize the
results. The scientific goals of this study were: 

1) To perform a detailed comparison of the ECMWF-analysis data with
radiosoundings and AWS-data (Automatic Weather Station) of the wind
speed and the temperature, near the surface as well as in the free
atmosphere. 

2) Using radiosoundings to estimate the median values of the wind
speed in the first tens of meters at the South Pole and Dome C. 

3) Studying the wind speed in the free atmosphere we intend to
quantify which site is the best for astronomical applications. 

4) We extend the analysis of the Richardson number done by Geissler \&
Masciardri (\cite{ges06}) at Dome C to the three sites (South Pole,
Dome C and Dome A) in order to quantify the regions and periods that
are more likely to trigger turbulence.

\section{Results and Discussion}

\subsection{$\partial\theta/\partial z$ and ($\partial v/\partial z)^2$ at
  Dome A, Dome C and the South Pole}

As shown by  Hagelin {\em et al.\/} (\cite{hag08a}) there is an
excellent agreement between the ECMWF analysis and the
radiosoundings at both Dome C and the South Pole. However, during the
winter, there is also a large offset in the first vertical
grid-point. 

The dynamic instabilities are described by the wind shear and the
thermal stability is represented by the gradient of the potential
temperature. A positive potential temperature gradient is defined as stable
conditions, the vertical displacement of the air is suppressed and so is the
production of dynamic turbulence.

The monthly median of the potential temperature, see Fig.~4 of
Hagelin {\em et al.\/} (\cite{hag08b}), shows that during most of the year
a temperature inversion is present. In the central months of the winter
(June, July and August) Dome A shows the most stable conditions with a
very sharp temperature inversion near the surface.

The difference in ($\partial v/\partial z)^2$ is much less evident,
see Fig.~5 in Hagelin {\em et al.\/} (\cite{hag08a}), but Dome A has a
slightly larger gradient than the other two sites during most months.

\subsection{Radiosoundings: the surface wind speed}

Above the summits of the Internal Antarctic Plateau the surface winds
are expected to be weaker than elsewhere on the plateau. Fig.~6 in
Hagelin {\em et al.\/} (\cite{hag08a}) shows the median wind speed at
the South Pole and Dome C from April to November. While it is true that
the wind speed at the lowest level is weaker at the summit (Dome C)
than at the slope (South Pole), it is clearly visible a sharp wind
shear in the first 10/20 m at Dome C. Above this height the wind speed
at Dome C is either stronger or comparable to that of the South Pole.

In the core of the winter (June, July and August) the wind speed above
Dome C reaches 8 m/s at 20 m and 9 m/s at 30 m. The sharp change in
the wind speed in the first 10/20 m matches our expectations of a
large wind speed gradient. This is a necessary condition to justify
the presence of optical turbulence in the surface layer (Agabi {\em et
  al.\/} \cite{Ag06}, Trinquet {\em et al.\/} \cite{Tr08}) in spite of
very stable thermal conditions. Such a strong wind speed, 8-9 m/s at
10 m, might be a source of vibrations produced by the impact of the
atmospheric flow on a telescope structure and should therefore be
taken into account in the design of astronomical facilities.

\subsection{ECMWF analyses: The wind speed in the free atmosphere}


The wind speed in the free atmosphere 
during the summer is quite weak and is almost constant with height
(see Fig.~7 in Hagelin {\em et al.\/} 
(\cite{hag08a})). 
The median wind speed from December to March never
exceeds 15 m/s at any height for any site. As the winter approaches
the wind speed above 10 km increases monotonically. The rate at which
the wind speed increases is not the same above the different sites.
The smallest increase rate is seen at the South Pole whereas the
largest increase rate is found above Dome C. These differences are far
from being negligible since at these heights the median wind speed at
Dome C is almost twice that of the other sites.

This effect can be explained by the polar vortex, that
forms above Antarctica during the winter. The increase rate of the
wind speed is proportional to the distance of the site to the centre
of the polar vortex. This is the reason why the wind speed above Dome C is
particularly large above 10 km in winter. At 15 km  a.s.l. the wind
speed of Dome C can easily be almost twice the wind speed of Dome A or
F and even thrice the wind speed of the South Pole. The wind speed at
the South Pole is the weakest of the four in all seasons and at all
heights.

\subsection{The Richardson number}

The Richardson number is an indicator of the stability of the
atmosphere, $Ri=\frac{g}{\theta}\frac{\partial \theta/\partial
  z}{(\partial v/\partial z)^2}$, where g is the gravitational
acceleration (9.8 m/s$^2$), $\theta$ is the potential temperature and
v is the horizontal wind speed. The atmosphere is considered to be
stable if the Richardson number is larger than the critical value,
typically 0.25. The smaller the Richardson number is the higher is the
probability to trigger turbulence.

Comparing the Richardson number gives a relative estimate of the
probability of the triggering of turbulence.  The method proposed by
Geissler \& Masciadri (\cite{ges06}), to calculate the inverse of the
Richardson number to rank different sites qualitatively, has been
definitely proved by comparing an Antarctic site (Dome C) with a
mid-latitude one (Mt.~Graham).  Indeed, as seen in Fig.~9 in Hagelin
{\em et al.\/} \cite{hag08a}, Dome C is always more stable than
Mt.~Graham, except the high part of the atmosphere in September and
October when the polar vortex creates strong high altitude winds at
Dome C. As discussed in Hagelin {\em et al.\/} \cite{hag08a} this
proves that the probability to trigger turbulence above mid-latitude
sites is larger than above Dome C.

Comparing the three Antarctic sites with each other, see Fig.~10 of
Hagelin {\em et al.\/} \cite{hag08a}, it appears that the most stable
conditions are found above the South Pole. Dome A is less stable than
the South Pole and the least stable conditions are found at Dome
C. This is most likey due to the polar vortex that produces a strong
wind shear at this site.

\section{Conclusions}
This study allowed us to draw a first comprehensive picture of the
atmospheric properties above the Internal Antarctic Plateau. In spite
of the generally good conditions for astronomical applications, Dome C
does not appear to be the best site with respect to the wind speed, in
the free atmosphere as well as in the surface layer. All the other
sites show a weaker wind speed in the free atmosphere. 

Above Dome A the gradient of the potential temperature is particularly
large near the surface, indicating extreme thermal stability
associated to a strong optical turbulence when a thermodynamic
instability occurs. It is even possible that the optical turbulence
here is larger than at Dome C. However, to predict the thickness of
such a layer it is necessary to have either measurements or
simulations with a mesoscale model with a higher spatial resolution. 

At present, the best argument that makes Dome C a better place for
astronomical applications than the South Pole is the extreme thinness
of the turbulent surface layer. Dome A probably has comparable or
larger values of the optical turbulence with respect to Dome C in the
surface layer. We cannot conclude whether the surface layer at Dome A
is thinner than what is observed at Dome C. Our study clearly
indicates that Dome C is not the best site on the Internal Antarctic
Plateau with respect to the wind speed (both in the surface layer and
in the free atmosphere) nor is it the site with the most stable
conditions in the free atmosphere. Both the South Pole and Dome A show
more stable values of 1/Ri.

\section*{Acknowledgments}
This study has been carried out using radiosoundings from the AMRC
(Antarctic Meteorological Research Center), University of Wisconsin,
Madison and from the Progetto di Ricerca 'Osservatorio Meteo
Climatologico' of the PNRA (Programma Nazionale di Ricerche in
Antartide). ECMWF products are extracted from the MARS-catalogue. This
study has been funded by the Marie Curie Excellence Grant (ForOT)
MEXT-CT 2005-023878, FP6.

\end{document}